\documentclass[journal,onecolumn]{IEEEtran}
\ifCLASSINFOpdf
\else
\fi
\begin{document}
%
\title{From Senseless Swarms to Smart Mobs:\\ Tuning Networks for Prosocial Behaviour}
%
%
%

\author{        Sun Sun Lim, 
Roland Bouffanais~\IEEEmembership{(Member,~IEEE)}
\thanks{S.S. Lim is Professor of Communication and Technology, Head of Humanities, Arts and Social Sciences, Singapore University of Technology and Design e-mail: sunsun\_lim@sutd.edu.sg. }
\thanks{R. Bouffanais is Associate Professor, Engineering Product Development Pillar, Singapore University of Technology and Design e-mail: bouffanais@sutd.edu.sg.}
}

\maketitle

\begin{abstract}
Social media have been seen to accelerate the spread of negative content such as disinformation and hate speech, often unleashing reckless herd mentality within networks, further aggravated by malicious entities using bots for amplification. So far, the response to this emerging global crisis has centred around social media platform companies making reactive moves that appear to have greater symbolic value than practical utility. These include taking down patently objectionable content or manually deactivating the accounts of bad actors, while leaving vast troves of negative content to circulate and perpetuate within social networks. Governments worldwide have thus sought to intervene using regulatory tools, with countries such as France, Germany and Singapore introducing laws to compel technology companies to take down or correct erroneous and harmful content. However, the relentless pace of technological progress enfeebles regulatory measures that seem fated for obsolescence.
\end{abstract}

\begin{IEEEkeywords}
Social networks, Swarm, Technology \& Society.
\end{IEEEkeywords}

%
\IEEEpeerreviewmaketitle

\vspace{6ex}
\IEEEPARstart{I}{n} view of this ceaselessly evolving technological climate, we believe that a more enduring solution to the spread of negative content via social networks is to consciously design and tune networks so as to advance prosocial behaviour. We therefore propose a more proactive approach of tapping local perturbations in social networks to trigger prosocial responses and positive action. By tuning and tweaking the topology of social networks to favour prosocial behaviour, we can more effectively contain the spread of negative and deleterious content and forge a more robust information landscape.

\section*{Smart mobs or senseless swarms?}
In his prescient book Smart Mobs published in 2002~\cite{1}, Howard Rheingold postulated that when people are connected by devices with both communication and computing capabilities, they can collaborate and coordinate action to effect positive transformations. He highlighted political demonstrations that toppled oppressive regimes, the emergence of Wikipedia and open source collaborative software development, collective citizen response to disasters, and citizen journalism. Rheingold was farsighted, predicting the likes of the Arab Spring protests of 2010, the Occupy Wall Street movement of 2011 or even apps to help refugees such as RefAid and REFUNITE. While social scientists such as Rheingold and Manuel Castells~\cite{1,2} have recognised the limitations of these networked social movements, they have nevertheless recognised their potential for empowerment.

And yet, even as the strength of networks and wisdom of crowds have been mobilised with salutary results, we have also witnessed their vast potential for negativity. Hitherto, it does seem as if the banes of social networks have been far more salient than their boons. For every Ice Bucket Challenge that successfully raises funds to battle disease, there are many risky social media trends such as the Tide Pod, A4 Waist or even Choking Game memes that can cause severe harm to individuals and society at large. Online falsehoods too have been empirically proven to disseminate further and faster than accurate information. Notably, an MIT study found that false rumours are 70\% more likely to be re-tweeted than true rumours~\cite{3}.  

Disturbingly, the intensifying use of social media platforms such as Facebook and Twitter, and of participatory media such as Reddit and YouTube, has created a potent conduit for the propagation of disinformation, online vigilantism and hate speech. Globally, such damaging content has been seen to gain rapid and widespread traction on social media platforms, with algorithms igniting their virality, and user interactions escalating the hostility. Consequently, rather than facilitate the healthy sharing of diverse perspectives, discussions in social media platforms often descend into reductionist arguments, obnoxious name-calling and close-minded grandstanding. Far from being smart mobs, such senseless swarms converge around disinformation and foment discord. Mobilised by falsehoods and toxic views, they are fractious, divisive and undermine the very bedrock of our society.

\section*{Programming for positivity}

In view of such adverse effects of social network interactions, how can social networks be redesigned or tuned to exploit and amplify their positive potential? Recent research at the intersections of collective decision-making, network science and control theory may offer some surprising insights and potential strategies.

Seeking to bridge knowledge from social science and data science to better understand human collective behaviour, Alex Pentland postulated the concept of `idea flow' in his book Social Physics~\cite{4}. Pentland conceives of `idea flow' as a collective dynamic process in which specific human collective behaviours are forged through a combination of two forces: social learning (a.k.a. social influence) and social pressure. Fads are prototypical short-lived forms of human collective behaviour and social norms are a longer-term example. 

Let us consider how social networks can be harnessed and influenced for the spread of positive behavior. In 2010, Damon Centola reported a ground-breaking experiment on the spread of behavior in an online social network~\cite{5}. It revealed the critical influence of the structure or topology of the underlying network on the effectiveness of the spreading process. Pentland conducted a similar experiment, reported in his Harvard Business Review article titled ``Beyond the Echo Chamber"~\cite{6}. It involved an online trading platform, eToro, that enabled online human traders to engage in social trades. The collective dynamics of these online traders indirectly interacting through eToro were consistent with the empirical result uncovered by Centola. 

Both studies found that the complex topology of the social network---how people are connected or interact---is critical to the effectiveness of the spreading process. When people are too interconnected, social exploration is greatly reduced and herding behaviour emerges, thus producing the echo chamber effect of the same views being perpetuated. On the other hand, when people are connected in a way that promotes the flow of ideas, social exploration is heightened and a diversity of ideas is shared, thereby allowing the wisdom of the crowd to peak. Departing from this premise, we must thus modulate social networks so as to ensure that people are optimally (and not excessively) connected such that they freely exchange ideas, and are liberated from the vicious recycling of trite concepts.

\section*{The potential of perturbations}

Beyond merely optimising the topology of social networks, we can also consider introducing stimuli or perturbations to trigger or elicit prosocial behaviour. By tweaking the topology of social networks, it is possible to slow down and even quarantine perturbations to impede the flow of negative views. At the same time, the spread of positive ideas can be promoted by triggering prosocial responses through the appropriate channelling of positive perturbations.

Drawing from the insights of network science and control theory, Roland Bouffanais and collaborators theoretically proved the existence of an optimal network topology for yielding the most effective collective response~\cite{7}. Their study was inspired by rich evidence of collective behaviours in the animal kingdom: swift evasive manoeuvres in flocks of starling being attacked by predators, cooperative transportation of cereal crumbs by groups of ants etc. Nature demonstrates that a certain level of mimicry of knowledgeable individuals, i.e. social learning, vastly improves the accuracy of collective decision-making that is critical to the survival of a group and species: foraging strategies, mate identification or habitat selection.

In their experiments, Bouffanais and collaborators introduced local perturbations into a swarm to investigate how group-level consensus was attained. They subjected an artificial swarm of robots to perturbations of varying speeds to study the phenomenon of mimicry and the effectiveness of their collective decision-making process. By tuning the interaction network between the swarming bots, it was possible to unfreeze the misguided herd behaviour (the swarm's echo chamber) and foster a significantly more responsive collective behaviour to fast-paced perturbations. In other words, if we extend this knowledge to the social networks we are currently familiar with, dynamically altering the extent of interaction between users can restore a rich flow of ideas, thereby preventing users from converging around points of disinformation and discord, and help to avert damaging herd mentality. 

Such experiments are revealing because they shed light on collective behaviours that are not naturally occurring in nature. Biologists are unable to tweak the interaction networks of flocking birds or schooling fish, and ethologists have no control over the pace of local perturbations affecting swarms of animals. Instead, experiments with artificial swarm robotics systems can inform the development of strategies for modulating idea flow in social networks so as to privilege positive idea flow while discouraging negative ones. Technologically-mediated social networks are ultimately manmade and subject to human intervention, and the algorithms that drive such social interaction are fundamentally perturbations. The question then remains as to how we can introduce carefully-calibrated perturbations that can yield prosocial responses to enhance the integrity of social networks. 

As promising as this proposed approach sounds, it is critical to bear in mind that social networks are `complex animals' with multiplex or multilayer networks. Data scientist Alain Barrat and his collaborators recently demonstrated that different layers of a social network do not always convey the same information~\cite{8}. Hence, the proposed approach of introducing perturbations may be challenging to implement in an online social network replete with human flaws and foibles, unlike a comparatively straightforward robotic swarm interaction network. Nonetheless, given that social media platforms collect copious amounts of information on their users, such data can be systematically mined to identify the most tweakable layer at which to introduce perturbations, and at what pace.

\section*{The ethical burden}

As Sheila Jasanoff has rightly cautioned, technologists can exult in the promise of their innovations, but must be cognizant of the unexpected harms and unintended consequences these may unleash~\cite{9}. It would appear that as currently designed, online social networks can wreak considerably more detriment than benefit. The insinuation of such sociotechnical architectures into our everyday interpersonal interactions can compound destructive human behaviour if their effects are not acknowledged or arrested. Ultimately, while regulatory responses stem from a groundswell of societal despair, they are no match for the tidal wave of challenges technology has ushered in. Technologists must therefore confront their lofty ethical burden because with power comes responsibility. Rather than rue the unforeseen adverse outcomes of their technological wizardry, they must press on to devise creative solutions to mitigate, if not obliterate, the problems that their innovations have introduced. After all, technologists have successfully plumbed their intellectual prowess to forge creative spaces for social interaction. So, too, should they tap their technological genius to shape senseless swarms into smart mobs.

\ifCLASSOPTIONcaptionsoff
  \newpage
\fi



%

\end{document}